\documentclass[conference]{IEEEtran}
\ifCLASSINFOpdf
\else
\fi
\usepackage{amsmath,dsfont,bbm,epsfig,amssymb,amsfonts,amstext,verbatim}\usepackage{amsopn,cite,subfigure}
\usepackage{multirow,multicol,lipsum,xfrac}
\usepackage{amsthm}
\usepackage{mathtools,amsthm}
\usepackage{perpage}
\usepackage{balance}
\usepackage{url}
\usepackage{amsfonts}
\usepackage{epsfig}
\usepackage[font={small}]{caption}
\usepackage{psfrag}	
\usepackage{etoolbox}
\usepackage{algorithmicx}
\usepackage[Algorithm,ruled]{algorithm}
\usepackage{algpseudocode}
\usepackage{pifont}
\usepackage[utf8]{inputenc}
\usepackage[T1]{fontenc}  
\usepackage[nolist]{acronym}
\usepackage{paralist}
\usepackage{enumitem}
\usepackage{bbm}
\usepackage[process=auto]{pstool}
\usepackage{tikz,pgfplots}
\usetikzlibrary{shapes,arrows,chains,matrix,positioning,scopes}

\newcommand{\setR}{\mathbb{R}}

\newcommand{\setC}{\mathbb{C}}

\newcommand{\brc}[1]{\left( #1 \right) }
\newcommand{\dbc}[1]{\left[ #1 \right] }
\newcommand{\prc}[1]{\mathrm{Prc}\left\lbrace #1 \right\rbrace }
\newcommand{\diag}[1]{\mathrm{diag}\left\lbrace #1 \right\rbrace }

\newcommand{\rmp}{\mathrm{p}}

\newcommand{\her}{\mathsf{H}}

\newcommand{\bh}{{\mathbf{h}}}

\newcommand{\bx}{{\boldsymbol{x}}}

\newcommand{\xx}{\mathrm{x}}

\newcommand{\set}[1]{\left\lbrace#1\right\rbrace}

\newcommand{\bz}{{\boldsymbol{z}}}

\newcommand{\bg}{{\mathbf{g}}}
\newcommand{\bs}{{\boldsymbol{s}}}

\newcommand{\bv}{{\boldsymbol{v}}}

\newcommand{\bww}{\mathbf{w}}

\newcommand{\dif}{\mathrm{d}}

\newcommand{\trp}{\mathsf{T}}

\newcommand{\mA}{\mathbf{A}}
\newcommand{\mW}{\mathbf{W}}

\newcommand{\mI}{\mathbf{I}}

\newcommand{\mG}{\mathbf{G}}
\newcommand{\mGamma}{\mathbf{\Gamma}}
\newcommand{\mQ}{\mathbf{Q}}

\newcommand{\mD}{\mathbf{D}}

\newcommand{\mX}{\mathbf{X}}

\newcommand{\mT}{\mathbf{T}}
\newcommand{\mH}{\mathbf{H}}

\newcommand{\Ex}[2]{\mathbbmss{E}_{#2} \left\lbrace #1 \right\rbrace  }

\newcommand{\rzf}{{\mathsf{rzf}}}
\newcommand{\srzf}{{\mathsf{srzf}}}

\newcommand{\sinr}{{\mathrm{SINR}}}
\newcommand{\esnr}{{\mathrm{ESNR}}}

\newcommand{\argmin}{\mathop{\mathrm{argmin}}}

\newcommand{\rss}{\mathrm{RSS}}

\newcommand{\norm}[1]{\lVert #1 \rVert}

\newcommand{\abs}[1]{\lvert #1 \rvert}
\newcommand{\tr}[1]{\mathrm{tr} \left\lbrace #1 \right\rbrace }

\newtheoremstyle{mystyle}
  {}
  {}
  {}
  {}
  {\bfseries}
  {:}
  { }
  {}

\theoremstyle{mystyle}

\newtheorem{remark}{Remark}

\algnewcommand\algorithmicLet{\textbf{Let}}
\algnewcommand\Let{\item[\algorithmicLet]}
\algnewcommand\algorithmicSet{\textbf{Set}}
\algnewcommand\Set{\item[\algorithmicSet]}

\algnewcommand\algorithmicInitiate{\textbf{Initiate}}
\algnewcommand\Initiate{\item[\algorithmicInitiate]}
\algnewcommand\algorithmicStart{\textbf{Begin}}
\algnewcommand\Begin{\item[\algorithmicStart]}
\algnewcommand\algorithmicEnd{\textbf{End}}
\algnewcommand\End{\item[\algorithmicEnd]}

\algnewcommand\algorithmicOutP{\textbf{Output:}}
\algnewcommand\Out{\item[\algorithmicOutP]}

\algnewcommand\algorithmicInP{\textbf{Input:}}
\algnewcommand\In{\item[\algorithmicInP]}

\newcounter{bar}

\begin{document}
\title{Secure Regularized Zero Forcing for Multiuser MIMOME Channels}

\author{
\IEEEauthorblockN{
Saba Asaad\IEEEauthorrefmark{1},
Ali Bereyhi\IEEEauthorrefmark{1},
Ralf R. M\"uller\IEEEauthorrefmark{1}, and
Rafael F. Schaefer\IEEEauthorrefmark{2}
}
\IEEEauthorblockA{
\IEEEauthorrefmark{1}Institute for Digital Communications, Friedrich-Alexander Universit\"at Erlangen-N\"urnberg, Germany \\
\IEEEauthorrefmark{2}Information Theory and Applications Chair, Technische Universit\"at Berlin, Germany\\
\{saba.asaad, ali.bereyhi, ralf.r.mueller\}@fau.de, rafael.schaefer@tu-berlin.de
}
\thanks{This work has been presented in the 2019 Asilomar Conference on Signals, Systems, and Computers. The link to the final version in the proceedings will be available later.}
}
\IEEEoverridecommandlockouts

\maketitle

\begin{abstract}
This paper proposes a new linear precoding scheme for downlink transmission in MIMOME channels, referred to as \textit{secure regularized zero forcing}. The scheme modifies regularized zero forcing precoding, such that the beamformers further~suppress the information leakage towards the eavesdroppers. The proposed scheme is characterized in the large-system limit, and a closed-form expression for the achievable ergodic secrecy rate per user is derived. Numerical investigations demonstrate high \textit{robustness against the quality of eavesdroppers' channel}.\vspace*{3mm}
\end{abstract}

\IEEEpeerreviewmaketitle

\section{Introduction}

Conventional precoding schemes in \ac{mimome} channels \cite{khisti2010secure} are often linear and independent of eavesdroppers' \ac{csi} \cite{geraci2012secrecy,geraci2013linear,geraci2013large}. The linearity constraint is mainly imposed for computational tractability. The independency~from eavesdroppers' \ac{csi} further follows the fact that even simple beamforming towards the legitimate receivers suppresses the signal at the eavesdroppers effectively when the density of these malicious terminals in the network is low. In the asymptotic regime, this latter behavior is referred to as~\textit{secrecy-for-}\\ \textit{free} \cite{kapetanovic2015physical,bereyhi2018robustness} indicating that in massive \ac{mimo}  wiretap settings \cite{zhu2014secure} with a fixed number~of eavesdroppers, the information leakage vanishes as the number of antennas grows large, by using simple linear precoders.

Despite the above justifications, taking eavesdroppers' \ac{csi} into account at the precoder can result in significant performance enhancement, specially when 
\begin{inparaenum}
\item the density of malicious and legitimate terminals in the network is moderate or high, and
\item the quality of signals received by eavesdroppers is comparable to that of legitimate users.
\end{inparaenum}
Such scenarios are likely to occur in current and next generations of mobile~net-works, due to the high number of mobile devices.

\subsection{Contributions}
In this paper, we propose a new linear precoding scheme for downlink transmission in \ac{mimome} channels. The precoder follows the least-squares based approach, developed in \cite{bereyhi2018glse,bereyhi2017asymptotics,bereyhi2017nonlinear}, and extends the \ac{rzf} precoding scheme \cite{peel2005vector} to \ac{mimo} systems with multiple eavesdroppers. To study the performance of the proposed scheme, we derive a closed-form expression for the achievable secrecy rate per user in the system when the number of~transmit~antennas, legitimate receivers and eavesdroppers grow large with fixed ratios. Our large-system analysis extends the earlier results in \cite{geraci2012secrecy,nguyen2008multiuser} to the larger scope, and depicts tight consistency with simulations. Numerical investigations show that in contrast to \ac{rzf} precoding, the proposed scheme is more robust against the quality of eavesdroppers' channel and~report~a significant performance gain when the quality of the channel~to~the~eavesdroppers is better that that of legitimate users.

\subsection{Notations}
Throughout the paper, scalars, vectors, and matrices~are~represented by~non-bold, bold lower case, and bold upper case letters, respectively. The real axis and complex plane are shown by $\setR$ and $\setC$, respectively. $\mH^{\her}$, $\mH^{*}$, and $\mH^{\trp}$ are the conjugate transpose, conjugate, and transpose of $\mH$, respectively. $\norm{\mH}_F$ further denotes the Frobenius norm of $\mH$. $\log\left(\cdot\right)$ is the binary logarithm, and $[x]^+ \coloneqq \max\{0, x\}$. Expectation is denoted by $\Ex{.}{}$, and $\mathcal{CN} (\eta, \sigma^2)$ represents the complex Gaussian distribution with mean $\eta$ and variance $\sigma^2$. For brevity, the set $\set{1,\ldots,N}$ is abbreviated by $[N]$.

\section{Problem Formulation}
\label{sec:sys}
We consider a Gaussian multiuser \ac{mimo} wiretap channel with a \ac{bs}, $K$ legitimate receivers and $J$ eavesdroppers. The \ac{bs} is equipped with a transmit array of size $M$, and the receiving terminals, i.e., the legitimate receivers and the eavesdroppers, are single-antenna. The system is assumed to perform in \ac{tdd} mode, and hence the uplink and downlink channels are reciprocal. The \ac{csi} is estimated in uplink training mode and is known at the \ac{bs}, as well as the receiving terminals.

\subsection{System Model}
The \ac{bs} intends to transmit messages $m_k \in \dbc{2^{N R_k}}$, for $k\in\dbc{K}$, confidentially to legitimate receivers $k$. To this end, $m_k$ is first encoded into a codeword of length $N$, i.e. $\left[ s_k\brc{1}, \ldots ,\right.$ $\left. s_k\brc{N} \right]$, and then transmitted within $N$ transmission intervals over the channel as follows: At time instant $n$, the \ac{bs} maps the encoded vector $\bs\brc{n} = \left[ s_1\brc{n}, \ldots, s_K\brc{n}\right]^\trp$ to the \textit{transmit vector} $\bx\brc{n}\in\setC^M$ via the precoder $\prc\cdot:\setC^K \mapsto \setC^M$,~i.e. $\bx\brc{n} = \prc{ \bs\brc{n} }$, and transmits it via the antenna array.

We assume that the channel experiences quasi-static fading, such that its coherence time interval is larger than $N$ transmission intervals. Let $\bh_k \in\setC^M$ contain coefficients of the uplink channel~between legitimate receiver $k$ and the \ac{bs}. Due to the channel reciprocity, the receive signal in interval $n$ reads
\begin{align}
y_k \brc{n} = \bh_k^\trp \bx\brc{n} + w_k\brc{n}
\end{align}
where $w_k\brc{n}$ is complex white Gaussian noise with zero mean and variance $\sigma_k^2$, i.e., $w_k\brc{n} \sim \mathcal{CN}\brc{0, \sigma_k^2}$. After~$N$~intervals, the receiver recovers $\hat{m}_k = \phi_k \brc{ \set{y_k\brc{n}} }$, where $\set{y_k\brc{n}} = \set{y_k\brc{1},\ldots,y_k\brc{N}}$ and $\phi_k\brc\cdot$ denotes the decoder.

For $j \in \dbc{J}$, the $j$-th eavesdropper observes
\begin{align}
z_j \brc{n} = \bg_j^\trp \bx\brc{n} + v_j\brc{n}
\end{align}
by overhearing the channel. Here, $\bg_j\in\setC^M$ denotes the channel from the $j$-th eavesdropper to the \ac{bs}, and $v_j\brc{n}$ is additive white Gaussian noise which reads $v_j\brc{n} \sim \mathcal{CN}\brc{0, \rho_j^2}$.

To guarantee secure transmission, we consider the \textit{worst-case} scenario in which the eavesdroppers are cooperating. In this case, the secrecy rate tuple $\brc{R_1,\ldots,R_K}$ is said to be achievable, if there exist a sequence of encoders and decoders, indexed by $N$, such that
\begin{subequations}
\begin{align}
&\lim_{N\uparrow \infty} \max_{k\in\dbc{K}} \set{\Pr\dbc{\phi_k \brc{ \set{y_k\brc{n}} } \neq m_k} } = 0 \label{eq:Const1}\\
&\lim_{N\uparrow \infty} \hspace*{-.5mm} \frac{1}{N} \mathrm{I} \brc{\mathcal{S}\brc{m_1,\ldots,m_K};\set{z_1\brc{n},\ldots,z_J\brc{n}}} \hspace*{-.5mm} = \hspace*{-.5mm} 0 \label{eq:Const2}
\end{align}
for all $\mathcal{S}\brc{m_1,\ldots,m_K} \subseteq \set{m_1,\ldots,m_K}$, where $\mathcal{S}\brc\cdot$ is a subset of user messages. The constraint in \eqref{eq:Const1} guarantees the reliability of transmissions towards the legitimate receivers. Moreover, \eqref{eq:Const2} indicates that the signals received~by~the~eavesdroppers leak no information about any subset of the transmit messages.
\end{subequations}

\subsection{Achievable Secrecy Rate with Linear Precoding}
For linear precoders, the transmit vector is written as 
\begin{align}
\bx\brc{n} = \mW \bs\brc{n}
\end{align}
for some precoding matrix $\mW = \dbc{ \bww_1 , \ldots , \bww_K }$ satisfying the transmit power constraint $\Ex{\tr{\bx^\her\bx}}{}/M \leq P$. The vector $\bww_k \in \setC^M$, for $k\in\dbc{K}$, is referred to as the \textit{precoding vector} of~legitimate receiver $k$. Without loss of generality, we assume that $\Ex{\bs\brc{n} \bs^\her\brc{n} }{} = \mI$.

Following the discussions in \cite{ekrem2011secrecy,geraci2012secrecy}, the secrecy rates
\begin{align}
R_k = \dbc{\log \left( \frac{ 1+ \sinr_k}{ 1+ \esnr_k }  \right)}^+
\label{eq:R_k}
\end{align}
are shown to be achievable for $k\in\dbc{K}$, where
\begin{subequations}
\begin{align}
\sinr_k &= \frac{ \abs{\bh^\trp_k \bww_k}^2  }{ \sigma_k^2 + \sum\limits_{j=1, j\neq k}^K \abs{\bh^\trp_k \bww_j}^2 },\\
\esnr_k &= \norm{\mathbf{K}_{\rm Eve} \mG \bww_k}^2
\end{align}
\end{subequations}
$\mG = \dbc{\bg_1,\ldots,\bg_J}^\trp$ and $\mathbf{K}_{\rm Eve} = \diag{1/\rho_1 , \ldots , 1/\rho_J}$. In the sequel, we consider the achievable rates in \eqref{eq:R_k} as the metric to quantify the secrecy performance of the system. 

The main goal of this study is to design an~effective~linear precoding scheme which takes into account the secrecy restrictions imposed by eavesdroppers. We address this objective~by~modifying \ac{rzf} precoding \cite{peel2005vector}, such that the information leakage to the eavesdroppers is~efficiently~suppressed at the precoding stage. For sake of brevity, we drop the time index, i.e. $n$, throughout the derivations in the remaining parts of this paper.

\section{Secure RZF Precoding}
In \ac{rzf} precoding, the precoding matrix reads 
\begin{align}
\mW_\rzf\brc{\zeta} =  \sqrt{\frac{P}{\beta_{\rzf} \brc{\zeta}   } } \left. \mA \brc{\zeta} \right.
\label{eq:RZF0}
\end{align}
for the shaping matrix 
\begin{align}
\mA \brc{\zeta} = \mH^\her \brc{\mH\mH^\her + \left. \zeta \right. \mI_K}^{-1} 
\label{eq:RZF}
\end{align}
tuned by the \textit{regularizer} $\zeta$ and the scaling factor
\begin{align}
\beta_\rzf \brc{\zeta} = \frac{1}{M} {\tr{\mA \brc{\zeta} \mA^\her\brc{\zeta}}}
\end{align}
which guarantees the satisfaction of the transmit power constraint.~Here, $\mH \hspace*{-.5mm} = \hspace*{-.5mm} \dbc{\bh_1,\ldots,\bh_K}^\trp$ denotes the vector downlink channel. The shaping matrix in \eqref{eq:RZF} performs regularized~channel inversion. At $\zeta = 0$, \ac{rzf} precoding reduces to the~zero-forcing scheme. In general, $\zeta$ is tuned such that~a~given~performance metric, e.g. ergodic sum rate, is optimized.

\subsection{Alternative Formulation of RZF Precoding}
\ac{rzf} precoding is alternatively observed as the~\ac{rls}~solution to the following linear regression problem: Find matrix $\mW$, such that the linear expansion $\mH\bx$ with $\bx = \mW \bs$ approximates an scaled version of $\bs$, i.e. $\psi \bs$~for some $\psi$, with minimum \ac{lse}, subject to $\Ex{\norm{\bx}^2}{} \leq P$. Following the method of \ac{rls}, the solution to this problem is given by minimizing 
\begin{align}
\rss\brc{\mW} \coloneqq \Ex{\norm{\mH\mW \bs - \psi \bs}^2}{\bs},
\end{align}
known as the \ac{rss}, penalized by the power constraint. In other words, $\mW$ is found by
\begin{subequations}
\begin{align}
\mW &= \argmin_{ \mX \in \setC^{M\times K} } \left. \rss\brc{\mX} + \zeta \left. \Ex{\norm{\mX\bs}^2}{\bs} \right. \right. \label{eq:RZF_RLSa} \\
&=  \argmin_{ \mX \in \setC^{M\times K} }  \tr{\mX^\her\brc{ \mH^\her \mH + \zeta \mI_M} \mX - 2 \psi \Re\set{\mH \mX}} \nonumber\\
&= \psi \left. \mA\brc{\zeta} \right.
\label{eq:RZF_RLS}
\end{align}
\end{subequations}
where $\zeta$ is a Lagrange multiplier. By considering the transmit power constraint, \eqref{eq:RZF_RLS} reduces to \eqref{eq:RZF0}. Extension of this~\ac{rls} based approach to other constraints leads to \ac{glse} precoding which has been proposed and studied in \cite{bereyhi2017asymptotics,bereyhi2017nonlinear,bereyhi2018glse}.

\subsection{RLS-Based Precoding with Security Constraints}
Following the \ac{rls} interpretation of \ac{rzf} precoding,~the~secrecy constraint can be further imposed at the transmit side by penalizing the \ac{rss} term. To illustrate this point, let
\begin{align}
f_{\rm Eve}\brc{\mW}:\setC^{M \times K} \mapsto \setR^+_0
\end{align}
quantify the information leakage when the linear precoder~$\mW$ is employed. Let $f_{\rm Eve}\brc\cdot$ be proportional~to~the~information leakage meaning that $f_{\rm Eve}\brc{\mW_1} \leq f_{\rm Eve}\brc{\mW_2}$ indicates that $\bs$ is estimated from the overheard signals in $\bz_1 = \mG\mW_1 \bs+\bv$ with higher error~probability~compared to $\bz_2 = \mG\mW_2 \bs+\bv$. 

Given $f_{\rm Eve}\brc\cdot$, a secrecy constraint can be imposed on the system by restricting the precoding matrix $\mW$ to satisfy
\begin{align}
f_{\rm Eve}\brc{\mW } \leq L
\end{align}
for some information leakage $L$. The \ac{rls} formulation in this case can be modified by penalizing the \ac{rss} term with~both~the power and secrecy constraints. That means $\mW$ is set to
\begin{align}
\mW  \hspace*{-1mm} = \hspace*{-1mm} \argmin_{ \mX \in \setC^{M\times K} }  \rss\brc{\mX} + \lambda \left. \Ex{\norm{\mX\bs}^2}{\bs} \right. + \theta f_{\rm Eve}\brc{\mX} 
\label{eq:Secure_RZF}
\end{align}
for some tunable factors $\lambda$ and $\theta$. The optimization in~\eqref{eq:Secure_RZF}~simultaneously reduces the \ac{lse} at the legitimate terminals and the leakage towards the eavesdroppers. 


Deriving a function which analytically characterizes~the~information leakage is not a tractable task. Nevertheless, one can consider an alternative metric which is proportional to the capability of the eavesdroppers in decoding the information. 
To find such a metric, we note that in the ideal case~with~significantly narrow beamforming, we desire to have 
\begin{align}
\abs{\bg_j^\trp\bww_k } = 0,
\end{align}
for $j\in [J ]$ and $k \in [K]$. This indicates that a natural choice for $f_{\rm Eve}\brc\cdot$ is 
\begin{subequations}
\begin{align}
f_{\rm Eve}\brc{\mW} &= \sum_{j=1}^J \sum_{k=1}^K \abs{\bg_j^\trp\bww_k }^2\\
&= \norm{\mG\mW}_{F}^2 = \tr{ \mW^\her \mG^\her \mG \mW}
\label{eq:f_Eve}
\end{align}
\end{subequations}
By substituting \eqref{eq:f_Eve}, \eqref{eq:Secure_RZF} reduces to a convex optimization problem whose solution is $\mW = \psi \mA\brc{\lambda,\theta}$ where
\begin{align}
\mA \brc{\lambda,\theta} = \brc{\mH^\her \mH + \left. \theta \right. \mG^\her \mG + \lambda \left. \mI_M \right.}^{-1} \mH^\her.
\end{align}
By restricting the transmit power to $P$, the \textit{\ac{srzf}} precoder is concluded as
\begin{align}
\mW_\srzf\brc{\lambda,\theta} =  \sqrt{\frac{P}{\beta_{\srzf}\brc{\lambda,\theta}}} \left. \mA \brc{\lambda,\theta} \right.
\label{eq:SRZF}
\end{align}
with 
\begin{align}
\beta_{\srzf} \brc{\lambda,\theta} = \frac{1}{M} \left. \tr{\mA \brc{\lambda,\theta} \mA^\her\brc{\lambda,\theta}} \right. .
\end{align}
Note that $\mW_\srzf\brc{\zeta,0} = \mW_\rzf\brc{\zeta}$. In fact, the \ac{srzf} scheme utilizes the \ac{csi} of the malicious terminals and modifies \ac{rzf} beamformers, such that leakage to the eavesdroppers is further suppressed. In general, $\theta$ and $\lambda$ are tuned such that a  given  performance metric, e.g. ergodic  sum rate,~is~optimized.

\section{Large-System Analysis}
In this section, the large-system performance of the proposed precoding scheme is characterized. To this end,~we~consider a scenario in which the number of transmit antennas~$M$, number of legitimate receivers $K$ and number of eavesdroppers $J$ are significantly large; however, the ratios
\begin{subequations}
\begin{align}
\alpha_{\rm l} &= \frac{K}{M}\\
\alpha_{\rm o} &= \frac{J}{M}
\end{align}
\end{subequations}
are constant. We refer to $\alpha_{\rm l}$ as the \textit{legitimate channel load}, and to $\alpha_{\rm o}$ as the \textit{overhearing channel load}. For sake of brevity, we further assume that
\begin{itemize}
\item For $k\in[K]$ and $j\in[J]$, $\bh_k$ and $\bg_j$ are are \ac{iid} Gaussian vectors with zero mean and variance $1/M$.
\item $\rho_j^2 = \rho^2$ for $j\in[J]$, and $\sigma_k^2 = \sigma^2$ for $k\in[K]$.
\end{itemize}

To start the derivations, let us define 
\begin{align}
\mQ &= \mH^\her \mH + \left. \theta \right. \mG^\her \mG + \lambda \left. \mI_M \right. .
\end{align}
Hence, the $k$-th beamformer of the \ac{srzf} precoder reads
\begin{align}
\bww_k = \sqrt{\frac{P}{\beta_{\srzf}\brc{\lambda,\theta}}} \left. \mQ^{-1} \bh_k^* \right. .
\end{align}
As a result, the \ac{sinr} at legitimate receiver $k$ reads
\begin{align}
\sinr_k = \dfrac{ \mu_{\rm l} U_k }{ \beta_{\srzf}\brc{\lambda,\theta}  + \mu_{\rm l} I_k }.
\end{align}
where $\mu_{\rm l} \coloneqq P/\sigma^2$ is the receive \ac{snr} at the legitimate terminals, and
\begin{align}
U_k &= \abs{\bh^\trp_k \mQ^{-1} \bh_k^*}^2, \\
I_k &= \sum\limits_{j=1, j\neq k}^K \abs{\bh^\trp_k \mQ^{-1} \bh^*_j}^2.
\end{align}
Using the Sherman-Morrison lemma, it is shown that
\begin{subequations}
\begin{align}
U_k &= \brc{\frac{\bh^\trp_k \mQ_k^{-1} \bh_k^*}{1+\bh^\trp_k \mQ_k^{-1} \bh_k^*}}^2 \\ 
I_k &= \sum\limits_{j=1, j\neq k}^K \frac{\abs{\bh^\trp_k \mQ_{k,j}^{-1} \bh_j^*}^2}{ 
\brc{1+\bh^\trp_k \mQ_k^{-1} \bh_k^*}^2 \brc{1+\bh^\trp_j \mQ_{k,j}^{-1} \bh_j^*}^2
} 
\end{align}
\end{subequations}
where $\mQ_k \coloneqq \mQ - \bh_k^* \bh_k^\trp$ and $\mQ_{k,j} \coloneqq \mQ - \bh_k^* \bh_k^\trp - \bh_j^* \bh_j^\trp$. For scaling factor $\beta_{\srzf}\brc{\lambda,\theta}$, we further can write
\begin{subequations}
\begin{align}
\beta_{\srzf} \brc{\lambda,\theta} &= \frac{1}{M} \left. \tr{\mA \brc{\lambda,\theta} \mA^\her\brc{\lambda,\theta}} \right. \\
&= \frac{1}{M} \tr{ \mH \mQ^{-2}\mH^\her  }\\
&= \frac{1}{M} \sum_{k=1}^K \bh_k^\trp \mQ^{-2} \bh_k^*\\
&\stackrel{\star}{=} \frac{1}{M} \sum_{k=1}^K \frac{\bh_k^\trp \mQ_k^{-2} \bh_k^*}{\brc{1+\bh_k^\trp \mQ_k^{-1} \bh_k^*}^2}
\end{align}
\end{subequations}
where $\star$ follows from the Sherman-Morrison lemma.

For $\esnr_k$, we can similarly write
\begin{align}
\esnr_k &= \frac{1}{\rho^2} \norm{\mG\bww_k}^2 = \frac{\mu_{\rm o} L_k}{\beta_{\srzf} \brc{\lambda,\theta}} 
\end{align}
where $\mu_{\rm o} \coloneqq P/\rho^2$ is the receive \ac{snr} at the eavesdroppers, and
\begin{subequations}
\begin{align}
L_k &= \bh_k^\trp \mQ^{-1} \mG^\her \mG \mQ^{-1} \bh_k^* \\
&= \frac{\bh_k^\trp \mQ_k^{-1} \mG^\her \mG \mQ_k^{-1} \bh_k^*}{\brc{1+\bh_k^\trp \mQ_k^{-1} \bh_k^*}^2}\\
&= \frac{1}{\brc{1+\bh_k^\trp \mQ_k^{-1} \bh_k^*}^2} \left.
\tr{ 
\mG \mQ_k^{-1} \bh_k^* \bh_k^\trp \mQ_k^{-1} \mG^\her 
}\right.\\
&= \frac{1}{\brc{1+\bh_k^\trp \mQ_k^{-1} \bh_k^*}^2} 
\sum_{j=1}^J  \left.
\abs{\bg_j \mQ_k^{-1} \bh_k^*}^2
\right.\\
&= \frac{1}{\brc{1+\bh_k^\trp \mQ_k^{-1} \bh_k^*}^2} 
\sum_{j=1}^J  \left.
\frac{\abs{\bg_j \mGamma_{k,j}^{-1} \bh_k^*}^2}{ \brc{1+ \theta \left. \bg_j \mGamma_{k,j}^{-1} \bg_j^*\right. }^2}
\right. .
\end{align}
\end{subequations}
Here, we define $\mGamma_{k,j} \coloneqq \mQ - \bh_k^* \bh_k^\trp - \theta \left. \bg_j^* \bg_j^\trp \right.$.

\subsection{Asymptotics via Free Probability}
To determine the asymptotic limits, we note that
\begin{enumerate}
\item Any two independent Hermitian random matrices, which are unitarily invariant, are asymptotically free \cite{tulino2004random}. This result indicates that
\begin{itemize}
\item $\bh_k^* \bh_k^\trp$ and $\mQ_k^{-1}$ are asymptotically free for $k\in[K]$.
\item $\bh_k^* \bh_k^\trp$ and $\mQ_{k,j}^{-1}$ are asymptotically free for $k , j \in[K]$.
\item $\bh_k^* \bh_k^\trp$ and $\mGamma_{k,j}^{-1}$ are asymptotically free for $k\in[K]$.
\item $\bg_j^* \bg_j^\trp$ and $\mGamma_{k,j}^{-1}$ are asymptotically free for $j\in[J]$.
\end{itemize}
\item $\mQ_k$, $\mQ_{k,j}$, and $\mGamma_{k,j}$ are single-rank perturbations of $\mQ$. As a result, the asymptotic distribution of their eigenvalues~is similar to that of $\mQ$. 
\item Defining the matrix $\mT \coloneqq \mH^\her \mH + \left. \theta \right. \mG^\her \mG$, we have
\begin{align}
\mQ &= \mT + \lambda \left. \mI_M \right. . 
\end{align}
Let us denote the asymptotic distribution of the eigenvalues of $\mT$ with $\rmp_\mT\brc{t}$. The eigenvalues of $\mQ$ are shifted versions of the eigenvalues of $\mT$ and are asymptotically distributed by $\rmp_\mT\brc{t-\lambda}$.
\end{enumerate}

Considering the above findings, we can write
\begin{subequations}
\begin{align}
\lim_{M\uparrow\infty} &\frac{\bh^\trp_k \mQ_k^{-1} \bh_k^*}{M} = \lim_{M\uparrow\infty} \frac{\tr{\mQ_k^{-1} \bh_k^* \bh^\trp_k  }}{M} \\
&\stackrel{\dagger}{=} \lim_{M\uparrow\infty}\frac{\tr{\mQ_k^{-1}}  }{M} \times \lim_{M\uparrow\infty}\frac{\tr{\bh_k^* \bh^\trp_k  }}{M} 
\end{align}
\end{subequations}
where $\dagger$ follows the fact that $\bh_k^* \bh_k^\trp$ and $\mQ_k^{-1}$ are asymptotically free. Noting that 
\begin{align}
\lim_{M\uparrow\infty} \tr{\bh_k^* \bh^\trp_k  }  =  \lim_{M\uparrow\infty} \norm{\bh_k}^2 = 1,
\end{align}
we have
\begin{subequations}
\begin{align}
\lim_{M\uparrow\infty} \bh^\trp_k \mQ_k^{-1} \bh_k^* &= \lim_{M\uparrow\infty} \frac{1}{M} \tr{\mQ_k^{-1}} \\
&= \Ex{\frac{1}{T+\lambda}}{T} = G_{\mT}\brc{-\lambda}
\end{align}
\end{subequations}
for some $T\sim\rmp_\mT$ with Stieltjes transform $G_{\mT}\brc{ \cdot }$ defined as
\begin{align}
G_{\mT}\brc{ s } = \int \frac{\rmp_\mT\brc{t} \dif t}{ t-s}.
\end{align}
Clearly, the limit does not depend on $k$. As a result,
\begin{align}
U_k = \brc{\frac{ G_{\mT}\brc{-\lambda}}{1+G_{\mT}\brc{-\lambda}}}^2
\end{align}
for $k\in[K]$. By same lines of derivations, we have
\begin{subequations}
\begin{align}
I_k &= \alpha_{\rm l} \brc{1+\Ex{\frac{1}{T+\lambda}}{T}}^{-4} \Ex{\frac{1}{\brc{T+\lambda}^2}}{T} \\
&= \dfrac{\displaystyle\alpha_{\rm l} G'_{\mT}\brc{-\lambda} }{ \displaystyle \brc{1+G_{\mT}\brc{-\lambda}}^4},
\end{align}
\end{subequations}
and
\begin{subequations}
\begin{align}
\beta_{\srzf} \brc{\lambda,\theta} &= \dfrac{\displaystyle\alpha_{\rm l} G'_{\mT}\brc{-\lambda} }{ \displaystyle \brc{1+G_{\mT}\brc{-\lambda}}^2}\\
L_k &= \dfrac{\alpha_{\rm o} G'_{\mT}\brc{-\lambda} }{ \brc{1+G_{\mT}\brc{-\lambda}}^2 \brc{1+ \theta G_{\mT}\brc{-\lambda}}^2}.
\end{align}
\end{subequations}
Consequently, $\sinr_k$ in the large-system limit converges to
\begin{align}
\sinr_{\rm asy} &= \dfrac{\displaystyle \mu_{\rm l} G^2_{\mT}\brc{-\lambda} \brc{1+G_{\mT}\brc{-\lambda}}^2 }{ \displaystyle \alpha_{\rm l} G'_{\mT}\brc{-\lambda} \left[ \mu_{\rm l} + \brc{1+G_{\mT}\brc{-\lambda}}^2 \right]} \label{eq:SINR_limit}
\end{align}
and the asymptotic limit of $\esnr_k$ is
\begin{align}
\esnr_{\rm asy} &= \dfrac{\displaystyle \mu_{\rm o} \alpha_{\rm o} }{ \displaystyle \alpha_{\rm l} \brc{1+ \theta G_{\mT}\brc{-\lambda}}^2 }. \label{eq:ESNR_limit}
\end{align}
To determine this limit for particular scenarios, we further need to determine the Stieltjes transform $G_{\mT}\brc{\cdot}$. 
\begin{remark}
Note that throughout the derivations, we did not utilize the Gaussianity of the channel matrices. In fact, the results in \eqref{eq:SINR_limit} and \eqref{eq:ESNR_limit} are valid for any unitarily invariant $\mH$ and $\mG$ whose row vectors are jointly independent.
\end{remark}

\subsection{Stieltjes Transform of $\mT$}
To find the Stieltjes Transform of $\mT$, we rewrite $\mT$ as
\begin{align}
\mT = \tilde{\mH}^\her \mD \tilde{\mH}
\end{align}
where $\tilde{\mH}$ is defined as
\begin{align}
\tilde{\mH} \coloneqq [\bh_1 , \ldots , \bh_K, \bg_1,\ldots,\bg_J]^\trp,
\end{align}
and $\mD \in \setR^{{(K+J)} \times {(K+J)}} $ is a diagonal matrix whose first $K$ diagonal entries are one and the rest are $\theta$. Noting that $\tilde\mH$ fulfills the conditions for the deformed quarter circle law, one can invoke the Silverstein-Bai result \cite{muller2013applications,silverstein1995empirical} and write
\begin{align}
G_{\mT}\brc{s} = \frac{1}{ \displaystyle -s + \brc{\alpha_{\rm l} + \alpha_{\rm o}} \int \frac{y \rmp_\mD\brc{y} \dif y}{1+ y G_{\mT}\brc{s}}} \label{eq:recurse}
\end{align}
where $\rmp_\mD\brc\cdot$ denotes the distributions of the diagonal entries of $\mD$ and reads
\begin{align}
\rmp_\mD\brc{y} = \frac{\alpha_{\rm l}}{\alpha_{\rm l} + \alpha_{\rm o}} \left. \delta\brc{y-1} \right. + \frac{\alpha_{\rm o}}{\alpha_{\rm l} + \alpha_{\rm o}} \left. \delta\brc{y-\theta} \right. .
\end{align}
Substituting into \eqref{eq:recurse}, we finally conclude
\begin{align}
1+ s G_{\mT}\brc{s} = \alpha_{\rm l} \frac{ G_{\mT}\brc{s} }{1+ G_{\mT}\brc{s}} + \alpha_{\rm o} \frac{ \theta G_{\mT}\brc{s} }{1+ \theta G_{\mT}\brc{s}} 
\end{align}
By taking derivative from the both sides of this equation, we further conclude
\begin{align}
G'_{\mT}\brc{s} = \frac{G_{\mT}\brc{s}}{ \displaystyle -s + 
\frac{ \alpha_{\rm l} }{\brc{1+ G_{\mT}\brc{s}}^2} + \frac{ \theta\alpha_{\rm o}  }{\brc{1+ \theta G_{\mT}\brc{s}}^2}
}.\label{eq:G_p}
\end{align}
As a result, $G_{\mT}\brc{-\lambda}$ is found as the positive solution of
\begin{align}
\lambda +  \frac{ \alpha_{\rm l} }{1+ \xx} + \frac{ \theta \alpha_{\rm o} }{1+ \theta \xx}  = \frac{1}{\xx}
\end{align}
and $G'_{\mT}\brc{-\lambda}$ is calculated by
\begin{align}
G'_{\mT}\brc{-\lambda} = \frac{ \xx }{ \displaystyle \lambda + 
\frac{ \alpha_{\rm l} }{\brc{1+ \xx}^2} + \frac{ \theta \alpha_{\rm o}  }{\brc{1+ \theta \xx}^2}}. 
\end{align}
By standard lines of derivations, one can show that the results recover the earlier derivations in \cite{geraci2012secrecy,nguyen2008multiuser}.

\section{Numerical Investigations}
\begin{figure}
\input{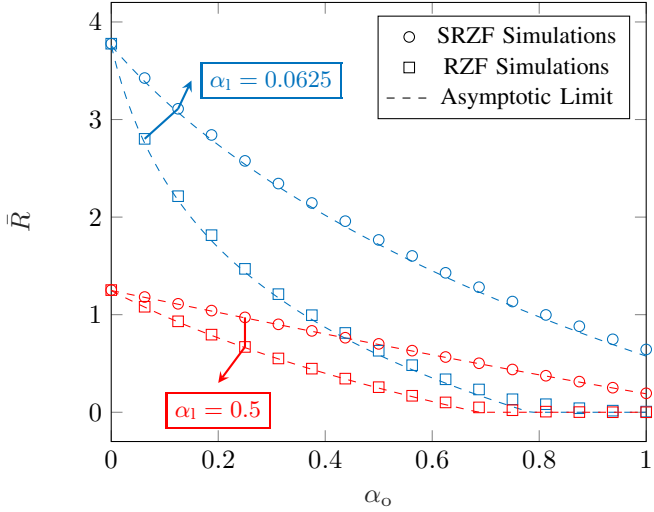}
\caption{Average ergodic secrecy rate vs. the overhearing channel load. The simulations are given for $M=128$ and $\log \mu_{\rm l} = \log \mu_{\rm o} = 0$ dB. For \ac{srzf} precoding, $\lambda= \theta = 1$, and $\zeta=1$ in \ac{rzf}.}
\label{fig:1}
\end{figure}

To investigate the performance of the proposed precoding scheme and validate the analytic derivations, we study sample scenarios via numerical simulations. As an overall measure of performance, we define the \textit{average ergodic secrecy rate} as
\begin{align}
\bar{R} = \frac{1}{K} \sum_{k=1}^K \Ex{R_k}{ }.
\end{align}
Following the asymptotic analysis, as $M$ grows large, $R_k \to \Ex{R_k}{ }$ and $\Ex{R_k}{ } \to \bar{R}$ for $k\in[K]$. Thus,
\begin{align}
\lim_{M\uparrow\infty} \bar{R} = \left[ \log\brc{\frac{1+\sinr_{\rm asy}}{1+\esnr_{\rm asy}}} \right]^+. \label{eq:LimitR}
\end{align}

To validate the large-system results, we sketch in Fig.~\ref{fig:1} the average ergodic secrecy rate against the overhearing channel load $\alpha_{\rm o}$ for two scenarios; namely, a scenario with \textit{high user density}, i.e. $\alpha_{\rm l} = 0.5$, and a scenario with \textit{low user density}, i.e. $\alpha_{\rm l} = 0.0625$. In both scenarios, the \ac{snr} at all receive terminals is set to one, i.e. $\log \mu_{\rm l} = \log \mu_{\rm o} = 0$ dB. For sake of comparison, the results are given for both the \ac{srzf} and \ac{rzf} schemes, where in the \ac{srzf} precoder $\lambda= \theta = 1$, and in \ac{rzf} $\zeta=1$. The entries of $\mH$ and $\mG$ are generated \ac{iid} with $\mathcal{CN}\brc{0,1/M}$. The figure shows the simulations for $M=128$ transmit antennas, as well as the results given via asymptotic analyses. It is seen that the analytic derivations are tightly consistent with the simulation results. From the figure, it is further observed that for the given setting the proposed scheme is constantly outperforming \ac{rzf}. This observation however needs further investigations, as in this setting the tunable parameters, i.e. $\lambda$ and $\theta$, are kept fixed.

\subsection{Optimal Tuning of SRZF Precoding}

\begin{figure}
\input{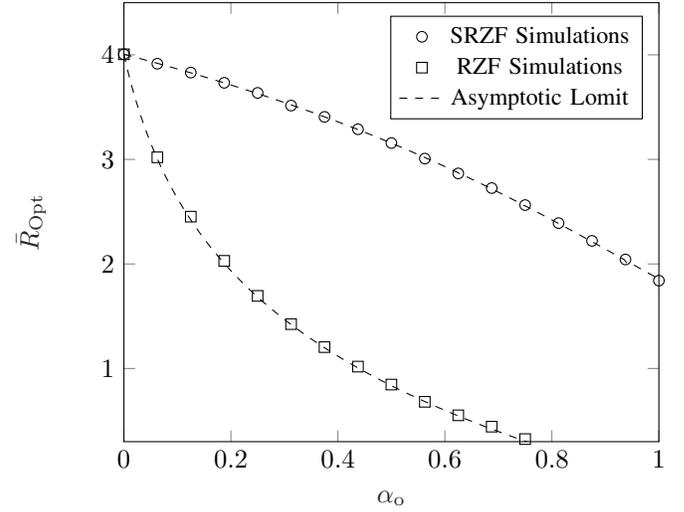}
\caption{Optimized average ergodic secrecy rate vs. the overhearing channel load. The simulations are given for $M=128$. The results are sketched for $\log \mu_{\rm l} = \log \mu_{\rm o} = 0$ dB and $\alpha_{\rm l} = 0.0625$.}
\label{fig:2}
\end{figure}
Following the tight consistency of analytical results, seen in Fig.~\ref{fig:1}, we tune the \ac{srzf} precoder by optimizing~the~asymptotic limit of the average ergodic secrecy rate. In other words, for a given setting, we find $\lambda$ and $\theta$ such that the limiting value of $\bar{R}$, given in \eqref{eq:LimitR}, is maximized. We denote this maximum value with $\bar{R}_{\rm Opt}$. 

Fig.~\ref{fig:2} shows $\bar{R}_{\rm Opt}$ versus the overhearing channel load for the low user density scenario in Fig.~\ref{fig:1}. The optimal choice of $\lambda$ and $\theta$, as well as the optimal \ac{rzf} regularizer, is found at each point by maximizing the asymptotic limit in \eqref{eq:LimitR}. The simulation points are then calculated by simulating the system with optimally tuned parameters for $M=128$. 

Considering Fig.~\ref{fig:2}, two findings are demonstrated:
\begin{inparaenum}
\item For non-zero overhearing channel loads, the \ac{srzf} scheme constantly outperforms \ac{rzf}. This finding is intuitive, since for $\alpha_{\rm o} \neq 0$, the so-called \textit{secrecy-for-free} property, reported in \cite{bereyhi2018robustness}, does not hold anymore, and hence, further suppression of leakage by \ac{srzf} enhances the performance.
\item The gap between the secrecy rates achieved by \ac{srzf} and \ac{rzf} increase as $\alpha_{\rm o}$ grows. This observation comes from the fact that as in networks with \textit{high density of eavesdroppers}, beamforming based on the legitimate channel information results in high  information leakage, and hence modification of the beamformers based on the eavesdropper channel improves the secrecy performance considerably.
\end{inparaenum}

\subsection{Robustness of SRZF Precoding}
To study the further gains proposed by the \ac{srzf} scheme, we plot the optimized average ergodic secrecy rate against the receive \ac{snr} at the eavesdroppers, i.e. $\mu_{\rm o}$, for the two scenarios, considered in Fig.~\ref{fig:1}, when $\alpha_{\rm o} = \alpha_{\rm l}/2$. In this figure, $\log \mu_{\rm l} = 0$ dB, and $\log \mu_{\rm o}$ is swept between~$-8$~and $8$ dB. The figure depicts that the average secrecy rate achieved by the \ac{rzf} scheme drops significantly, as the receive \ac{snr} at the eavesdroppers increases. The \ac{srzf} scheme however is very robust. This phenomena demonstrates the efficiency of  beamforming modification proposed by \ac{srzf} and indicates its \textit{robustness against the quality of eavesdroppers' channels}.

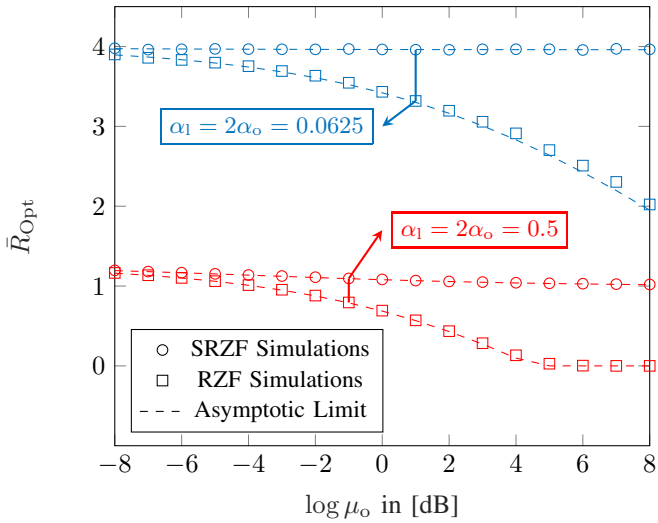
\begin{figure}
%
%
\definecolor{mycolor1}{rgb}{0.00000,0.44706,0.74118}%
\begin{tikzpicture}

\begin{axis}[%
width=2.8in,
height=2.3in,
at={(1.97in,1.037in)},
scale only axis,
xmin=-8,
xmax=8,
xtick={-8,-6,-4,-2,0,2,4,6,8},
xticklabels={{$-8$},{$-6$},{$-4$},{$-2$},{$0$},{$2$},{$4$},{$6$},{$8$}},
xlabel style={font=\color{white!15!black}},
xlabel={$\log \mu_{\rm o}$ in [dB]},
ymin=-1,
ymax=4.5,
ytick={0,1,2,3,4},
yticklabels={{$0$},{$1$},{$2$},{$3$},{$4$}},
ylabel style={font=\color{white!15!black}},
ylabel={$\bar{R}_{\rm Opt}$},
axis background/.style={fill=white},
legend pos= south west]
\addlegendimage{only marks, mark=o}
\addlegendimage{only marks, mark=square}
\addlegendimage{no marks, dashed}
\addplot [color=red, draw=none, mark=o, mark options={solid, red}]
  table[row sep=crcr]{%
-8	1.19638964744964\\
-7	1.18155477417829\\
-6	1.16749172561783\\
-5	1.15277697432078\\
-4	1.13829371400933\\
-3	1.12553297403821\\
-2	1.11073744523848\\
-1	1.0955843482108\\
0	1.08278373178749\\
1	1.06841866848209\\
2	1.05738875740505\\
3	1.04976630741207\\
4	1.04151332449634\\
5	1.03413252285789\\
6	1.02966847024406\\
7	1.02386490820606\\
8	1.01954707464509\\
};
\addlegendentry{\small{SRZF Simulations}}

\addplot [color=red, draw=none, mark=square, mark options={solid, red}]
  table[row sep=crcr]{%
-8	1.1643004019973\\
-7	1.13546395500002\\
-6	1.10212391024109\\
-5	1.06101869683766\\
-4	1.01192116806252\\
-3	0.95392416360578\\
-2	0.880819547440119\\
-1	0.794266431619318\\
0	0.692344572567626\\
1	0.570453200927221\\
2	0.435001642881286\\
3	0.283930628185319\\
4	0.132699769194444\\
5	0.0272531438989303\\
6	0.00329929798533713\\
7	8.7553576580062e-05\\
8	2.68186399439737e-05\\
};
\addlegendentry{\small{RZF Simulations}}

\addplot [color=red, solid, thick, forget plot]
  table[row sep=crcr]{%
-1	0.794266431619318\\
-1	1.0955843482108\\
};

\addplot [color=red, -stealth, thick, forget plot]
  table[row sep=crcr]{%
-1	1.0955843482108\\
0	1.7\\
};
\node[right,red] at (axis cs:0,1.7) {\fbox{\small{$ \alpha_{\rm l} = 2 \alpha_{\rm o} =  0.5$}}};

\addplot [color=red, dashed]
  table[row sep=crcr]{%
-8	1.19351146841644\\
-7	1.180510225479\\
-6	1.16653870779023\\
-5	1.15185449010686\\
-4	1.13677560941271\\
-3	1.1216565476241\\
-2	1.10685762106274\\
-1	1.09271278495776\\
0	1.07950150949399\\
1	1.06742947857068\\
2	1.0566206268772\\
3	1.04712025183774\\
4	1.03890660993639\\
5	1.03190723273174\\
6	1.02601628534149\\
7	1.02111023978368\\
8	1.01706037327444\\
};
\addplot [color=red, dashed, forget plot]
  table[row sep=crcr]{%
-8	1.16153134916841\\
-7	1.13436049626665\\
-6	1.10086634759981\\
-5	1.05977617559373\\
-4	1.00965716944435\\
-3	0.948940244479213\\
-2	0.875965444779822\\
-1	0.78905180682725\\
0	0.686590802442456\\
1	0.567156970193022\\
2	0.429623166942745\\
3	0.273263074583341\\
4	0.0978225489818647\\
5	0\\
6	0\\
7	0\\
8	0\\
};
\addlegendentry{\small{Asymptotic Limit}}
\addplot [color=mycolor1, draw=none, mark=o, mark options={solid, mycolor1}, forget plot]
  table[row sep=crcr]{%
-8	3.97857121265898\\
-7	3.9624920766518\\
-6	3.96544951027973\\
-5	3.97082613779613\\
-4	3.9674468334224\\
-3	3.96551673143655\\
-2	3.96470230083798\\
-1	3.96950742785733\\
0	3.9635658611735\\
1	3.96002607821973\\
2	3.95940215404123\\
3	3.9655775154965\\
4	3.9664595039849\\
5	3.96625244132879\\
6	3.9566565238035\\
7	3.97133762544248\\
8	3.96488597404856\\
};
\addplot [color=mycolor1, draw=none, mark=square, mark options={solid, mycolor1}, forget plot]
  table[row sep=crcr]{%
-8	3.90130839143754\\
-7	3.85942439144994\\
-6	3.83269976576012\\
-5	3.79828655894613\\
-4	3.754241117437\\
-3	3.69309436067487\\
-2	3.63465027313159\\
-1	3.54620255176206\\
0	3.43304264631681\\
1	3.31843757576712\\
2	3.19576947616896\\
3	3.05790940673293\\
4	2.9125508311891\\
5	2.70308728874862\\
6	2.50867066623319\\
7	2.30428295999829\\
8	2.02194196214405\\
};

\addplot [color=mycolor1, solid, thick, forget plot]
  table[row sep=crcr]{%
1	3.96002607821973\\
1	3.31843757576712\\
};

\addplot [color=mycolor1, -stealth, thick, forget plot]
  table[row sep=crcr]{%
1	3.31843757576712\\
0	3\\
};
\node[left,mycolor1] at (axis cs:0,3) {\fbox{\small{$ \alpha_{\rm l} = 2 \alpha_{\rm o} =  0.0625$}}};

\addplot [color=mycolor1, dashed, forget plot]
  table[row sep=crcr]{%
-8	3.97329455712942\\
-7	3.97126475880767\\
-6	3.96945733955299\\
-5	3.96787709315877\\
-4	3.96651748852149\\
-3	3.96536385877827\\
-2	3.96439652529566\\
-1	3.96359345377053\\
0	3.96293226957391\\
1	3.96239162854014\\
2	3.9619520356322\\
3	3.96159624065856\\
4	3.96130933897105\\
5	3.96107868477151\\
6	3.96089369867597\\
7	3.96074562652892\\
8	3.96062728639179\\
};
\addplot [color=mycolor1, dashed, forget plot]
  table[row sep=crcr]{%
-8	3.8955808365705\\
-7	3.86840998366875\\
-6	3.83491583500189\\
-5	3.79382566299582\\
-4	3.74370665684644\\
-3	3.68298973188131\\
-2	3.61001493218192\\
-1	3.52310129422935\\
0	3.42064028984455\\
1	3.30120645759511\\
2	3.16367265434483\\
3	3.00731256198544\\
4	2.83187203638396\\
5	2.63759504971984\\
6	2.42519877675587\\
7	2.19580306108305\\
8	1.95082836951035\\
};
\end{axis}
\end{tikzpicture}%
\caption{Optimized average ergodic secrecy rate vs. the receive \ac{snr} at the eavesdroppers. The simulations are given for $M=128$ and $\log \mu_{\rm l} = 0$ dB.}
\label{fig:3}
\end{figure}

\section{Conclusions}
A novel linear precoding scheme has been proposed for downlink transmission in \ac{mimome} channels. A closed-form expression for the asymptotic achievable secrecy rate~per~user has been derived for this precoder. The large-system results depict tight consistency with simulations, and hence can be employed to tune the precoder. Numerical investigations have demonstrated high performance enhancements achieved by the proposed scheme. Specifically, the precoder has shown to be highly robust against the change in the channel quality and outperform significantly the well-known \ac{rzf} scheme when the eavesdroppers experience better channel quality compared to the legitimate receivers.

\bibliography{ref}
\bibliographystyle{IEEEtran}

\begin{acronym}
\acro{mimo}[MIMO]{multiple-input multiple-output}
\acro{mimome}[MIMOME]{multiple-input multiple-output multiple-eavesdropper}
\acro{csi}[CSI]{channel state information}
\acro{awgn}[AWGN]{additive white Gaussian noise}
\acro{iid}[i.i.d.]{independent and identically distributed}
\acro{ut}[UT]{user terminal}
\acro{bs}[BS]{base station}
\acro{mt}[MT]{mobile terminal}
\acro{eve}[Eve]{eavesdropper}
\acro{lse}[LSE]{least squared error}
\acro{glse}[GLSE]{generalized least squared error}
\acro{rls}[RLS]{regularized least-squares}
\acro{rhs}[r.h.s.]{right hand side}
\acro{lhs}[l.h.s.]{left hand side}
\acro{wrt}[w.r.t.]{with respect to}
\acro{tdd}[TDD]{time-division duplexing}
\acro{papr}[PAPR]{peak-to-average power ratio}
\acro{mrt}[MRT]{maximum ratio transmission}
\acro{zf}[ZF]{zero forcing}
\acro{rzf}[RZF]{regularized zero forcing}
\acro{srzf}[SRZF]{secure \ac{rzf}}
\acro{snr}[SNR]{signal to noise ratio}
\acro{sinr}[SINR]{signal to interference plus noise ratio}
\acro{rf}[RF]{radio frequency}
\acro{mf}[MF]{match filtering}
\acro{mmse}[MMSE]{minimum mean squared error}
\acro{rss}[RSS]{residual sum of squares}
\acro{dc}[DC]{difference of convex functions}
\acro{dca}[DCA]{DC programming algorithm}
\end{acronym}

\end{document}